\documentclass[11pt]{article}
\newcommand{\be}{\begin{equation}}
\newcommand{\ee}{\end{equation}}
\newcommand{\bea}{\begin{eqnarray}}
\newcommand{\eea}{\end{eqnarray}}
\begin{document}

\title{Elements of Environmental Decoherence\thanks{
To be published in the proceedings of the Bielefeld conference on
``Decoherence: Theoretical, Experimental, and Conceptual Problems",
edited by P. Blanchard, D. Giulini, E. Joos, C. Kiefer, and I.-O.
Stamatescu (Springer 1999).
}}
%
%
\author{Erich Joos\\
{ Rosenweg 2, D-22869 Schenefeld, Germany}}

\date{}

\maketitle

\begin{abstract}

In this contribution I give an introduction to the essential
concepts and mechanisms of decoherence by the environment. The
emphasis will be not so much on technical details but rather
on conceptual issues and the impact on the interpretation
problem of quantum theory.
\end{abstract}

\section{What is decoherence?}

Decoherence is the irreversible formation of quantum correlations
of a system with its environment. These correlations lead to
entirely new properties and behavior compared to that shown by
isolated objects.

 Whenever we have a product state
of two interacting systems - a very special state - the unitary evolution
according to the Schr\"odinger equation will lead to entanglement,

\bea 
|\varphi\rangle|\Phi\rangle &
    \stackrel{t}\longrightarrow &
    \sum_{n,m}\limits c_{nm}|\varphi_n\rangle
    |\Phi_m\rangle\ \nonumber
\\
& \qquad = &\sum_n\limits \sqrt{p_n(t)} |\tilde{\varphi}_n(t) \rangle |\tilde{\Phi}_n(t) \rangle .
\eea
The rhs of (1) can no longer be written as a single product in the
general case. This can also be described by using the Schmidt representation, 
shown in the second line, where the presence of more than
one component is equivalent to the existence of quantum correlations.

If  many degrees of freedom are involved in this process, this 
entanglement will become practically irreversible, except for
very special situations. Decoherence is thus a quite normal and,
moreover, ubiquitous, 
quantum mechanical process. Historically, the important observation
was that this de-separation of quantum states
happens extremely fast for macroscopic objects \cite{Zeh70}.
The natural environment cannot simply be ignored or treated as a 
classical background in this case.

Equation (1) shows that there is an intimate connection to the theory of
 irreversible processes. However,
decoherence must not be identified or confused with dissipation: decoherence
precedes dissipation by acting on a much faster timescale, while
requiring initial conditions which are essentially the same as those 
responsible for the thermodynamic arrow of time \cite {ZehBook}.

When we consider observations at one of the two systems, we see various
consequences of this entanglement. First of all, our considered subsystem
will no longer obey a Schr\"odinger equation, the local dynamics is
in general very complicated, but can often be approximated by some sort
of master equation (The Schmidt decomposition
is directly related to the subsystem density matrices). The most important effect is the 
disappearance of
phase relations (i.e., interference) between certain subspaces of
the Hilbert space of the system. Hence the resulting superselection rules
can be understood as emerging from a dynamical, approximate and time-directed process.
If the coupling to the environment is very strong, the internal
dynamics of the system may become slowed down or even frozen. This is
now usually called the quantum Zeno effect, which apparently does not
occur in
our macroscopic world.

The details of the dynamics depend on the kind of coupling between the system we consider
and its environment. In many cases -- especially in the macroscopic domain --
this coupling leads to an evolution similar to a measurement process. Therefore it is
appropriate to recall the essential elements of the quantum theory 
of measurement.

\subsection{Dynamical Description of Measurement}

The standard description of measurement was laid down by von Neumann 
already in 1932 \cite{Neu32}. Consider a set of system states
$|n\rangle$ which our apparatus is built to discriminate. 
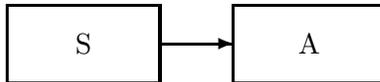
\begin{figure}[h]
\begin{center}
\setlength{\unitlength}{1cm}
\begin{picture}(5,1) \thicklines 
\put(0,0){\framebox(2,1){S}}
\put(3,0){\framebox(2,1){A}}
\put(2,0.5){\vector(1,0){1}}
\end{picture}
\end{center}
{Original form of the von Neumann measurement model. Information
about the state of the measured system S is transferred to the measuring
apparatus A.}

\end{figure}

For each state $|n\rangle$
we have a corresponding pointer state $|\Phi_n\rangle$ (more precisely, for
each ``quantum number" $n$ there exists a large set of macrostates
 $|\Phi_n^{(\alpha)}\rangle$, $\alpha$ describing microscopic degrees
of freedom). If the measurement is repeatable
 or ideal the dynamics of the measurement interaction must look like
\be |n\rangle|\Phi_0\rangle \quad \stackrel{t}{\longrightarrow} \quad
      |n\rangle|\Phi_n(t)\rangle\ . \ee
{}From linearity we can immediately see what happens for a general initial
state of the measured system,

\be \left(\sum_n c_n|n\rangle\right)|\Phi_0\rangle \quad
    \stackrel{t}\longrightarrow \quad \sum_n c_n|n\rangle
    |\Phi_n(t)\rangle\ . \ee

We do not find a certain measurement result, but a superposition. 
Through unitary evolution, a correlated (and still pure) state results,
 which contains
all possible results as components. Of course such a superposition must
not be interpreted as an ensemble. The transition from this
superposition to a single component -- which is what we observe --
constitutes the quantum measurement problem. As long as there is no collapse
we have to deal with the whole superposition -- and it is well known that
a superposition has very different properties compared to any of
its components. Quantum correlations are often misinterpreted as
(quantum) noise. This is wrong, however: Noise would mean that the considered system is
in a certain state, which may be unknown and/or evolve in a complicated way. Such
an interpretation is untenable and contradicts all experiments which
show the nonlocal features of quantum-correlated (entangled) states.

Von Neumann's treatment, as described so far, is unrealistic since
it does not take into account the essential openness of macroscopic
objects. This deficiency can easily be remedied by extending the
above scheme.

\subsection{Classical Properties through Decoherence}

If one takes into account that the apparatus A is coupled to its environment E, which
 also acts like a measurement device, the phase relations are (extremely fast) further
dislocalized into the total 
\begin{figure}[ht]
\begin{center}
\setlength{\unitlength}{1cm}
\begin{picture}(10,2)(-0.5,-0.5) \thicklines 
\put(0,0){\framebox(2,1){S}}
\put(3,0){\framebox(2,1){A}}
\put(2,0.5){\vector(1,0){1}}
\put(6,-0.5){\framebox(3,2){E}}
\put(5,0.3){\vector(1,0){1}}
\put(5,0.5){\vector(1,0){1}}
\put(5,0.7){\vector(1,0){1}}
\put(-0.5,-0.5){\dashbox{0.2}(6,2){}}
\end{picture}
\end{center}
{\footnotesize Realistic extension of the von Neumann
 measurement model. Information
about the state of the measured system S 
is transferred to the measuring
apparatus A and then very rapidly sent to the environment E.
The back-reaction on the (local) system S+A originates entirely from
quantum nonlocality.}

\end{figure}
system -- finally the entire universe,
according to
\be \left( \sum_n c_n|n\rangle|\Phi_n\rangle\right)|E_0 \rangle
  \quad  \stackrel{t}\longrightarrow \quad
   \sum_n c_n|n\rangle |\Phi_n\rangle |E_n\rangle .
\ee
The behavior of system+apparatus is then described by the density
matrix
\be
\rho_{SA} \approx \sum_n |c_n|^2 |n\rangle\langle n|
   \otimes |\Phi_n \rangle \langle \Phi_n |
   \qquad\mbox{if}\qquad
   \langle E_n|E_m \rangle \approx \delta_{nm}
\ee
which is identical to that of an ensemble of measurement results
$|n\rangle |\Phi_n\rangle$.

Of course, this does not resolve the measurement problem! This
density matrix describes only an ``improper" ensemble, i.e., with respect
to all possible observations at S+A it {\em appears} that a certain measurement
result has been achieved. Again, classical notions like noise or 
recoil are not appropriate: A acts dynamically on E, but the back-action
arises entirely from quantum nonlocality (as long as the measurement is 
``ideal", that is, (4) is a good approximation). Nevertheless, the system S+A
acquires classical behavior, since interference terms are absent with
respect to local observations if the above process is irreversible
\cite{Zur81,JZ85}.

\begin{samepage}
Needless to say, the interference terms still exist globally in the
total (pure) state, although they are unobservable at either system alone -- 
a situation which may be characterized by the statement

\begin{center}
{\em The interference terms still exist, but they are not {\em there}.}\cite{JZ85}
\end{center}
\end{samepage}

\section{Do we need observables?}

In most treatments of quantum mechanics the notion of an
observable plays a central role. Do observables represent a fundamental 
concept or can they
be derived? If we describe a measurement as a certain kind of 
interaction, then  observables should not be required as an
essential ingredient of quantum theory. In a sense
this was also done by von Neumann, but not used later very much
because of restrictions enforced by the Copenhagen school (e.g., 
the demand to describe a measurement device in
classical terms instead of seeking for a consistent treatment in terms of
wave functions).

Two elements are necessary to derive an observable that 
discriminates certain (orthogonal) system states $|n\rangle$.
First, one needs an appropriate interaction which is diagonal in the
eigenstates of the measured ``observable" and is able to ``move the pointer",
so that we have as above
\be
|n\rangle |\Phi_0\rangle 
\quad \stackrel{H_{int}}\longrightarrow \quad
|n\rangle |\Phi_n\rangle\ .
\label{meas}
\ee
This can be achieved by Hamiltonians of the form
\be 
H_{int} =\sum_n|n\rangle\langle n| \otimes\hat{A}_n\  
\ee
with appropriate $\hat{A}_n$ leading to orthogonal pointer states
(Note that (\ref{meas}) defines only the eigenbasis of an observable; the
eigenvalues represent merely scale factors and are therefore of
minor importance).
The second condition that must be fulfilled is dynamical stability
of pointer states against decoherence, that is, the pointer states
must only be passively recognized by the environment according to,
\be
|\Phi_n\rangle |E_0\rangle 
\quad \stackrel{decoherence}\longrightarrow\quad
 |\Phi_n\rangle |E_n\rangle\ . \label{dec}
\ee
{\em Both} conditions must be fulfilled. For example, a measurement device
which acts according to (\ref{meas}) would be totally useless, if it were not
stable against decoherence: Consider a Schr\"odinger cat state as
pointer state! The {\em same} basis states $|\Phi_n \rangle$ must be 
distinguished as dynamically relevant in (\ref{meas}) as well
as in (\ref{dec}).%
\renewcommand{\thefootnote}{\fnsymbol{footnote}}%
\footnote[1]{This explains {\em dynamically} why certain observables may
``not exist" operationally. For a general discussion of the relation between
quantum states and observables see Sect. 2.2 of \cite{Giu96}. Arguments along
these lines lead to the conclusion that one should not attribute a fundamental
status to the Heisenberg picture
 -- contrary to
widespread belief -- despite its {\em phenomenological} equivalence
with the Schr\"odinger picture.}%
\renewcommand{\thefootnote}{\arabic{footnote}}%

\section{Do we need superselection rules?}

What is a superselection rule? One way to define a superselection
rule is to say, that certain states $|\Psi_1 \rangle$, $|\Psi_2 \rangle$
are found in nature, but never general superpositions
$|\Psi \rangle = \alpha |\Psi_1 \rangle + \beta |\Psi_2 \rangle $. This
means that all observations can be described by a density matrix of the
form $\rho = p_1 |\Psi_1 \rangle \langle \Psi_1 | +
p_2 |\Psi_2 \rangle \langle \Psi_2 |$ . Clearly such a density matrix is
exactly what is obtained through decoherence in appropriate situations.

\subsection{Approximate superselection rules}

There are many examples, where it is hard to find certain superpositions in the 
real world. The most famous example has been given by Schr\"odinger: A 
superposition of a dead and an alive cat
\be |\Psi \rangle = |\mbox{dead cat} \rangle + |\mbox{alive cat} \rangle\  \ee
 is never observed, contrary to
what should be possible according to the superposition principle 
(and, in fact, {\em must} necessarily occur according to the
Schr\"odinger equation). Another drastic situation is given by a state
like
\be |\Psi \rangle = |\mbox{cat} \rangle + |\mbox{dog} \rangle\  .\ee
Such a superposition looks truly absurd, but  only  because we never
observe states of this kind! (The obvious objection that
one cannot superpose states of ``different systems" seems to
be inappropriate. For example, nobody hesitates to superpose
states with different numbers of particles.) A more down-to-earth example is given by the 
position of large objects, which are never found in states
\be |\Psi \rangle = |\mbox{here} \rangle + |\mbox{there} \rangle\  \label{loc} ,\ee
with ``here" and ``there" macroscopically distinct. Under realistic
circumstances such objects
are always well described by a localized
density matrix \mbox{$\rho (x,x') \approx p(x) \delta (x-x')$}. A special case
of this localization occurs in molecules (except the very small ones),
 which show a well-defined
spatial structure. The Born-Oppenheimer
approximation is not sufficient to explain this fact.

Quite generally we have an approximate superselection rule whenever we
describe the dynamics of a dynamical variable by some  rate
equation (that is, without interference) instead of the Schr\"odinger equation.

\subsection{Exact superselection rules}

Strict absence of interference can only be expected for discrete quantities.
One important example is electric charge. Can this be understood via
decoherence?
We know from Maxwell's theory, that every charge carries with itself an
associated electric field, so that a superposition of charges may
be written in the form \cite{Zeh}
\bea
\sum_q c_q |\Psi_q^{total}\rangle &=& 
   \sum_q c_q |\chi_q^{bare}\rangle | \Psi_q^{field}\rangle \nonumber \\
  & =& \sum_q c_q |\chi_q^{local}\rangle | \Psi_q^{far field}\rangle  \ .
\eea
Since we can only observe the local dressed charge, it has
to be described by the density matrix
\be
\rho= \sum_q |c_q|^2 |\chi_q^{local} \rangle\langle\chi_q^{local}|
\ee
{\em If} the far fields are orthogonal (distinguishable), coherence would be
absent locally. So the question arises: Is the Coulomb field only 
 part of the kinematics (implemented via the Gauss constraint) or
does it represent a quantum dynamical degree of freedom so that
 we have to consider decoherence via a retarded Coulomb field?
 For an attempt to understand part of the Coulomb field
as dynamical see \cite{Giu}.

What do experiments tell us? A superposition of the form (\ref{loc}) can be
observed for charged particles (cf. the contribution by Hasselbach\cite{Has}).
 On the other hand, the classical
(retarded) Coulomb field would contain information about the path of
the charged particle, destroying coherence. The situation does not appear very clear-cut.
Hence one essential question remains:
\begin{center}
\fbox
{ What is the {\em quantum} physical role of the Coulomb field?}
\end{center}

\noindent 
A similar situation arises in quantum gravity, where we can expect that
superpositions of different masses (energies) are decohered by the
spatial curvature.

Another important ``exact" superselection rule forbids superposing
states with integer and half-integer spin, for example
\be |\Psi \rangle = | \hbox{ spin 1} \rangle + | \hbox{ spin} \,{\scriptstyle ^1/_2} \rangle \ , \ee
which would  transform under a rotation by $2\pi $ into
\be |\Psi_{2\pi} \rangle = | \hbox{ spin 1} \rangle - | \hbox{ spin} \,{\scriptstyle ^1/_2} \rangle \ , \ee
clearly a different state because of the different relative phase. If one
{\em demands} that such a rotation should not change anything, such a
state must be excluded. This is one standard argument in favor of
the ``univalence" super\-selection rule.
On the other hand, one {\em has} observed the sign-change of spin\,${\scriptstyle ^1/_2}$ particles
under  a (relative) rotation by $2\pi$ in {\em certain} experiments. Hence we are left with two
options: Either we view the group SO(3) as the proper rotation group also in
quantum theory. Then nothing must change if we rotate the system by an angle of
$2\pi$. Hence  we can derive this superselection rule from symmetry. But this
may merely be a classical prejudice. The other choice is to use SU(2)
instead of SO(3) as rotation group. Then we are in need of explaining why those strange
superpositions never occur. This last choice amounts to keeping the
superposition principle as the fundamental principle of quantum theory. In more
technical terms we should then avoid using groups with non-unique (``ray"
\renewcommand{\thefootnote}{\fnsymbol{footnote}}%
\footnote[5]{
The widely used argument that physical states are to be represented by rays,
not vectors, in Hilbert space because the phase
of a state vector cannot be observed,
is misleading. Since relative phases are certainly relevant,
one should prefer  
a vector as a {\em fundamental} physical state concept,  rather than  a ray.
Rays cannot even be superposed without (implicitly) using vectors.
}%
\renewcommand{\thefootnote}{\arabic{footnote}}%
)
 representations, such as SO(3).
In supersymmetric theories, bosons and fermions are treated on an equal
footing, so it would be natural to superpose their states (what is
apparently never done in particle theory).

In a similar manner one could undermine the well-known argument leading
from the Galilean symmetry of nonrelativistic quantum mechanics to
the mass superselection rule. In this case we could maintain the
superposition principle and replace the Galilei group by a larger group. How
this can be done is shown by Domenico Giulini\cite{Giu}.

The final open question for this section then is: 

\begin{center}
\fbox
{Can {\em all} superselection rules be
understood as decoherence effects?}
\end{center}

\section{Examples}

\subsection{Localization}

The by now standard example of decoherence is the localization of
macroscopic objects.
Why do macroscopic objects always appear localized in space? Coherence 
between macroscopically different positions is destroyed {\it very}
rapid\-ly
 because
of the strong influence of scattering processes. The formal description
may proceed as follows. Let $|x\rangle$ be the position eigenstate
of a macroscopic object, and $|\chi\rangle$ the state of the
incoming particle.
 Following the von Neumann scheme (2), the scattering of
such particles off an object located at position $x$ may be written as
\be |x\rangle|\chi\rangle \stackrel{t}{\longrightarrow}
    |x\rangle|\chi_x\rangle=|x\rangle S_x|\chi\rangle\ , \ee
where the scattered state may conveniently be calculated by means of
an appropriate S-matrix. For the more general initial state of a wave
packet we have then
\be \int d^3x\ \varphi(x)|x\rangle|\chi\rangle
    \stackrel{t}{\longrightarrow}\int d^3x\ \varphi(x)|x\rangle
     S_x|\chi\rangle\ . \ee
Therefore, the reduced density matrix describing our object changes
into
\be \rho(x,x')=\varphi(x)\varphi^*(x')
  \left\langle\chi|S_{x'}^{\dagger}S_x|\chi\right\rangle\ .\label{ortho} \ee
Of course, a single scattering process will usually not resolve a small
distance, so
in most cases the matrix element on the right-hand side
 of (\ref{ortho}) will be close to unity.
If we add the contributions of many scattering processes, an
exponential damping of spatial coherence results:
\be \rho(x,x',t)= \rho(x,x',0)\exp\left\{-\Lambda t(x-x')^2\right\}\ .
\ee
The strength of this effect is described by a single parameter $\Lambda$
that may be called ``localization rate". It is given by
\be \Lambda= \frac{k^2Nv\sigma_{eff}}{V}\ .\ee
Here, $k$ is the wave number of the incoming particles, $Nv/V$
the flux, and $\sigma_{eff}$ is of the order of the total cross section
(for details see \cite{JZ85} or
Sect. 3.2.1 and Appendix 1 of \cite{Giu96}). 
Some values of $\Lambda$ are given in the table.

\begin{table}[htb]
{{\bf Localization rate} $\Lambda$ in $\mbox{cm}^{-2}
\mbox{s}^{-1}$ for three sizes of ``dust particles" and various
types of scattering processes (from \cite{JZ85}).
This quantity measures how fast interference between different
positions disappears as a function of distance in the course of
time.}
\begin{flushleft}
\renewcommand{\arraystretch}{1.2}  
\begin{tabular}{l|lll}
\hline
{} & \ $a=10^{-3}\mbox{cm}$ &\  $a=10^{-5}\mbox{cm}$ &\
$a=10^{-6}\mbox{cm}$\\
{} & \ dust particle        &\  dust particle        &\ large molecule
\\ \hline
Cosmic background radiation &\ $10^{6}$& $10^{-6}$ &$10^{-12}$ \\
300 K photons &\ $10^{19}$ & $10^{12}$ & $10^6$ \\
Sunlight (on earth) &\ $10^{21}$  & $10^{17}$ & $10^{13}$ \\
Air molecules & \  $10^{36}$ & $10^{32}$ & $10^{30}$ \\
Laboratory vacuum & \ $10^{23}$ & $10^{19}$ & $10^{17}$\\
($10^3$ particles/$\mbox{cm}^3$) & {}&{}&\\
\hline
\end{tabular}
\renewcommand{\arraystretch}{1}
\end{flushleft}\end{table}

Most of the numbers in the table are quite large, showing the extremely
strong coupling of macroscopic objects, such as dust particles, to their
natural environment. Even in intergalactic space, the 3K background
radiation
cannot simply be neglected.
\begin{samepage}

\noindent
Hence the main lesson is:
\medskip
\begin{center}
\fbox{{\bf Macroscopic objects are not even approximately isolated.}}
\end{center}
\medskip
\end{samepage}
\noindent A consistent unitary description must therefore include the environment
and finally the whole universe.%
\renewcommand{\thefootnote}{\fnsymbol{footnote}}%
\footnote[1]{One of the first stressing the importance of the
dynamical coupling of macro-objects to their environment was Dieter Zeh,
who wrote in his 1970 Found. Phys. paper \cite{Zeh70}: ``Since the interactions
between macroscopic systems are effective even at astronomical distances,
the only `closed system' is the universe as a whole. ... It is of course very
questionable to describe the universe by a wavefunction that obeys a
Schr\"odinger equation. Otherwise, however, there is no inconsistency
in measurement, as there is no theory."

This is now more or less commonplace, but this was not the case some 30 years ago,
when he sent an earlier version of this paper to the journal Il Nuovo Cimento. I quote
from the referee's reply: ``The paper is completely senseless. It is clear
that the author has not fully understood the problem and the previous
contributions in this field."  (H.D. Zeh, private communication)
}%
\renewcommand{\thefootnote}{\arabic{footnote}}%

If we combine this damping of coherence with the ``free" Schr\"odinger dynamics
we arrive at an equation of motion for the density matrix that to a good
approximation simply adds these two contributions,
\be i\frac{\partial\rho}{\partial t}
   =\left. [H_{internal},\rho]+i\frac{\partial\rho}{\partial t}
    \right\vert_{scatt.} \ . \ee
In the position representation this equation reads in one space dimension
\be i\frac{\partial\rho(x,x',t)}{\partial t}=
  \frac{1}{2m}\left(\frac{\partial^2}{\partial x'^2}
  -\frac{\partial^2}{\partial x^2}\right)\rho
  -i\Lambda(x-x')^2\rho\ . \ee
Solutions of this equation can easily be found (see, e.g.\cite{Giu96})

So far this treatment represents {\em pure} decoherence, following directly the von Neumann scheme. 
If recoil is added as a next step, we arrive at models including friction,
that is, quantum Brownian motion. 
There are several models
for
the quantum analogue of Brownian motion, some of which are even older
than
the first decoherence studies. Early treatments did not, however, 
draw a distinction
between decoherence and friction (decoherence alone does {\em not} imply friction.). As an example, consider the equation
of motion derived by Caldeira and Leggett \cite{CL83},
\be i\frac{\partial\rho}{\partial t}= [H,\rho]
   +\frac{\gamma}{2}[x,\{p,\rho\}]-i m\gamma k_BT[x,[x,\rho]] \ee
which reads for a ``free" particle
\bea i\frac{\partial\rho(x,x',t)}{\partial t}&=& \left[
  \frac{1}{2m}\left(\frac{\partial^2}{\partial x'^2}
  -\frac{\partial^2}{\partial x^2}\right)-i\Lambda(x-x')^2\right.
   \nonumber\\
   & & \left. +i\gamma(x-x')\left(\frac{\partial}{\partial x'}
   -\frac{\partial}{\partial x}\right)\right]
  \rho(x,x',t)\ , \eea
where $\gamma$ is the damping constant, and here $\Lambda=m\gamma k_BT$.

If one compares the effectiveness of the two terms representing
decoherence and relaxation, one finds that their ratio is given by
\be \frac{\mbox{decoherence rate}}{\mbox{relaxation rate}}
   =mk_BT(\delta x)^2\propto \left(\frac{\delta x}{\lambda_{th}}
     \right)^2 \ , \ee
where $\lambda_{th}$ denotes the thermal de~Broglie wavelength of the
considered object.
This ratio has for a typical macroscopic situation 
($m=1 \mbox{g}$, $T=300\mbox{K}$, $\delta x=1 \mbox{cm}$) the enormous
value of
about $10^{40}$! This shows that in these cases decoherence is {\em far
more important} than dissipation.

Not only the center-of-mass position of dust particles becomes
``classical" via 
decoherence. The spatial structure of molecules represents another most
important example. Consider a simple model of a chiral molecule.

Right- and left-handed versions both have a rather well-defined spatial
structure, whereas the ground state is -- for symmetry reasons --
a superposition of both chiral states. These chiral configurations are
usually separated by a tunneling barrier, which is so
high that under
normal circumstances tunneling is very improbable, as was already
shown by Hund in 1929. But this alone does not explain why chiral (and, indeed, most)
molecules are never found in energy eigenstates!

In a simplified model with low-lying nearly-degenerate eigenstates
$|1\rangle$
and $|2\rangle$, the right- and left-handed configurations may be given
by
\bea |L\rangle &=& \frac{1}{\sqrt{2}}(|1\rangle+|2\rangle)\nonumber\\
     |R\rangle &=& \frac{1}{\sqrt{2}}(|1\rangle-|2\rangle)\ . \eea
Because the environment recognizes the spatial structure via scattering
processes, only chiral
states are stable against decoherence,

\be |R,L\rangle|\Phi_0\rangle \stackrel{t}{\longrightarrow}
    |R,L\rangle|\Phi_{R,L}\rangle\ . \ee
The dynamical instability of energy (i.e., parity) eigenstates of molecules
represents a typical example of ``spontaneous symmetry breaking"
induced by decoherence.
Additionally, transitions between spatially oriented states are
suppressed by the quantum Zeno effect, described below.

\subsection{Quantum Zeno Effect}
The most dramatic consequence of a strong measurement-like interaction
of
a system with its environment is the quantum Zeno effect. It has been
discovered
several times and is also sometimes called ``watchdog effect" or
``watched pot
behavior",
although most people now use the term Zeno effect.
It is surprising only if one sticks to a classical picture
where observing a system and just verifying its state should have no
influence on it. Such a prejudice is certainly formed by our
everyday experience, where observing things in our surroundings does
not change their properties. As is known since the early times of quantum
theory, observation can drastically change the observed system.

The essence of the Zeno effect can easily be shown as follows. Consider
the
``decay" of a system which is initially prepared in the ``undecayed"
state $|u\rangle$. The probability to find the system undecayed, i.e.,
in the
same state $|u\rangle$ at time $t$ is for small time intervals given by
\bea P(t) &=& |\langle u|\exp(-i Ht)|u\rangle|^2 \nonumber\\
          &=& 1-(\Delta H)^2t^2+ {\cal O}(t^4) \label{zeno}\eea
with
\be (\Delta H)^2= \langle u|H^2|u\rangle - \langle u|H|u\rangle^2 \ .\ee
If we consider the case of $N$ measurements in the interval $ [0,t]$,
the
non-decay probability is given by
\be P_N(t)\approx\left[1-(\Delta H)^2\left(\frac{t}{N}\right)^2
    \right]^N > 1-(\Delta H)^2t^2= P(t)\ . \ee
This is always larger than the single-measurement probability
given by (\ref{zeno}).
In the limit of arbitrary dense measurements, the system no longer
decays,
\be P_N(t)= 1-(\Delta H)^2\frac{t^2}{N}+\ldots
    \stackrel{N\to\infty}{\longrightarrow}1\ .\ee
Hence we find that repeated measurements can completely hinder the
natural
evolution of a quantum system. Such a result is clearly quite distinct
from what is observed for classical systems. Indeed, the paradigmatic
example
for a classical stochastic process, exponential decay,
\be P(t)=\exp(-\Gamma t)\ , \ee
is not influenced
by repeated observations, since for $N$ measurements we simply have
\be P_N(t)=\left(\exp\left(-\Gamma\frac{t}{N}\right)\right)^N
          =\exp(-\Gamma t)\ . \ee

So far we have treated the measurement process in our discussion of the
Zeno
effect in the usual way by assuming a collapse of the system state onto
the subspace corresponding to the measurement result. Such a treatment
can be
extended by employing a von Neumann model for the measurement process,
e.g., by coupling a pointer to a two-state system. A simple toy model is
given
by the Hamiltonian
\bea H&=& H_0+H_{int}\nonumber\\
   &=& V(|1\rangle\langle 2|+|2\rangle\langle 1|)
   +E|2\rangle\langle 2|+\gamma\hat{p}(|1\rangle\langle 1|
   -|2\rangle\langle 2|)\ , \eea
where transitions between states $|1\rangle$ and $|2\rangle$ (induced by
the ``perturbation" {\it V}) are monitored by a pointer (coupling
constant $\gamma$). This model already shows all the typical features
mentioned above.

The transition probability starts for small times always quadratically,
according to the general result (\ref{zeno}). For times, where the pointer 
resolves the two states, a behavior similar to that found for Markow
processes appears: The quadratic time-dependence changes to a linear
one.
 For strong coupling the transitions are suppressed. This
clearly shows the dynamical origin of the Zeno effect.

An extension of the above model allows an analysis of the transition
from
the Zeno effect to master behavior (described by transition {\em rates} as was
first studied in quantum mechanics by Pauli in 1928). It can be shown that for many (micro-)states
 which are not sufficiently resolved by the environment, Fermi's
Golden
Rule can be recovered, with transition rates which are no longer
reduced by the Zeno effect. Nevertheless, interference between
macrostates
is suppressed very rapidly \cite{Jo84}.

\subsection{Decoherence of Fields}

In QED we find two (related) situations,
\begin{itemize}
\item ``Measurement" of charges by fields;
\item ``Measurement" of fields by charges.
\end{itemize}
In both cases, the entanglement between charge and field states leads
to decoherence as already described above in the discussion of
superselection rules, see also \cite{Giu96} and references therein.

In recent quantum optics experiments it is possible to prepare
and study superpositions of different classical field states,
quantum-mechanically represented by coherent states, for example
Schr\"odinger cat states of the form
\be
 |\Psi \rangle = N (|\alpha  \rangle + | -\alpha \rangle )
\ee
which can be realized as field states in a cavity. In these experiments
(see \cite{Brune96}) decoherence can be turned on gradually by coupling the
cavity to a reservoir. Typical decoherence times are in the range
of about 100 $\mu s$.

For {\em true} cats the decoherence time is much shorter (in particular,
it is {\em very much} shorter than the lifetime of a cat!). This leads
to the appearance of {\em quantum jumps}, although all underlying processes
are smooth in principle since they are governed by the
Schr\"odinger equation.

In experimental situations of this kind we find a gradual transition
from a superposition of different decay times (seen in ``collapse and
revival" experiments) to a local mixture of decay times (leading to
``quantum jumps") according to the following scheme.
\begin{figure}[h]
\begin{center}
\setlength{\unitlength}{1cm}
\begin{picture}(8,3) \thicklines 
\put(0,0){\makebox(4,1){\parbox{4cm}{local mixture of different decay times}}}
\put(6,0){\makebox(4,1){\parbox{4cm}{quantum jumps}}}
\put(0,1.6){\makebox(4,1){\parbox{4cm}{superposition of different decay times}}}
\put(6,1.6){\makebox(4,1){\parbox{4cm}{collapse and revivals}}}
\put(1,2.5){\makebox(2,1){\parbox{2cm}{theory}}}
\put(6.5,2.5){\makebox(2,1){\parbox{2cm}{experiment}}}
\put(1,0.8){\makebox(1,1){$\Downarrow$}}
\put(7,0.8){\makebox(1,1){$\Downarrow$}}
\put(0,2.7){\line(1,0){9}}
\put(5,0){\line(0,1){3.2}}
\end{picture}\end{center}
\end{figure}

\subsection{Spacetime and Quantum Gravity}

In quantum theories of the gravitational field, no classical
spacetime exists at the most fundamental level. Since it is 
generally assumed that the gravitational field has to be quantized,
the question again arises how the corresponding classical
properties can be understood. 

Genuine quantum effects of gravity are expected to occur for scales
of the order of the Planck length $\sqrt{G\hbar/c^3}$. It is
therefore often argued that the spacetime structure at larger scales
is automatically classical. However, this Planck scale argument
is as insufficient as the large mass argument in the evolution
of free wave packets. As long as the superposition principle is
valid (and even superstring theory leaves this untouched),
superpositions
of different metrics should occur at any scale. 

The central problem can already be demonstrated in a simple
Newtonian model\cite{Jo86}. Consider a cube of length
$L$ containing a homogeneous gravitational field with a
quantum state $\psi$ such that at some initial time $t=0$
\be  \vert\psi\rangle =c_1\vert g\rangle +c_2\vert g'\rangle
    \;  , 
\ee
where $g$ and $g'$ correspond to two different field strengths.
A particle with mass $m$
in a state $\vert\chi\rangle$, which moves through this volume,
``measures" the value of $g$, since its trajectory depends on the
acceleration $g$:
\be \vert \psi\rangle\vert\chi^{(0)}\rangle \to
    c_1\vert g\rangle\vert\chi_g(t)\rangle
    +c_2|g'\rangle|\chi_{g'}(t)\rangle\ .
  \ee
This correlation destroys the coherence between $g$ and $g'$,
and the reduced density matrix can be estimated to assume the
following form after many such interactions are taken into
account:
\be \rho(g,g',t) =\rho(g,g',0)\exp\left(-\Gamma t(g-g')^2\right),
       \ee
 where
 \[ \Gamma =nL^4\left(\frac{\pi m}{2k_{{B}}T}\right)^{3/2} \]
 for a gas with particle density $n$ and temperature $T$.
 For example, air under ordinary conditions, $L=1\mbox{ cm}$,  and
 $t=1\mbox{ s}$ yields a remaining coherence width of
 $\Delta g/g\approx 10^{-6}$\cite{Jo86}. 

Thus, matter does not only tell space to curve but also to
behave classically. This is also true in full quantum gravity.

In a fully quantized theory of gravity, for example in the canonical
approach described by the Wheeler-deWitt equation,
\be H |\Psi (\Phi,^{(3)}{\cal G}) \rangle =0 \ ,
\ee
where $\Phi$ describes matter and $^{(3)}{\cal G}$ is the three-metric,
everything is contained in the ``wave function of the universe" $\Psi$.
Here we encounter new problems: There is neither an external time parameter,
nor is there an external observer. How these problems can be
tackled is described in Claus Kiefer's contribution\cite{Kie}.

\section{Lessons}

What insights can be drawn from decoherence studies? It should be
emphasized that decoherence derives from a straightforward application of
standard quantum theory to realistic situations. It seems to be a
historical accident, that the importance of the interaction with
the natural environment was overlooked for such a long time.
Certainly the still prevailing (partly philosophical) attitudes enforced by the
Copenhagen school played a (negative) role here, for example by
outlawing a physical analysis of the measurement process in quantum-mechanical
terms.

Because of the strong
coupling of macroscopic objects, a quantum description of
macroscopic objects {\em requires} the inclusion of the natural
environment. A fully unitary quantum theory is only consistent
if applied to the whole universe. This does not preclude local phenomenological
descriptions. However, their derivation from a universal quantum theory and the
interpretation assigned to such descriptions have to be analyzed very
carefully.

We have seen that typical classical properties, such as
localization in space, are {\em created} by the environment in an
irreversible process, and are therefore not inherent attributes of macroscopic
objects. The features of the interaction define {\em what} is classical by
selecting a certain basis in Hilbert space.
Hence superselection sectors emerge from the dynamics. In all ``classical"
situations, the relevant decoherence time is extremely short, so that
the smooth Schr\"odinger dynamics leads to apparent discontinuities like
 ``events", ``particles"
or ``quantum jumps".

There are certain ironies in this situation.
{\em Local} classical properties find their explanation in
the {\em nonlocal} features of quantum states. Usually quantum objects are 
considered as fragile and easy to disturb, whereas macroscopic objects
 are viewed as
the rock-solid building  blocks of empirical reality. However, the opposite is
true: macroscopic objects are extremely sensitive and immediately
decohered.

On the practical side, decoherence also has its disadvantages. It
makes testing alternative theories difficult (more on that below), and
it represents a major obstacle for people trying to construct a quantum
computer. Building a really big one may well turn out to be as
difficult as detecting other Everett worlds!

\subsection{Does decoherence solve the measurement problem?}

Clearly not. What decoherence tells us, is that certain objects
{\em appear} classical when they are observed. But what is an
observation? At some stage, we still have to apply the usual probability
rules of quantum theory. These are hidden in density 
matrices, for example.

\subsection{Which interpretations make sense?}

One could also ask: what interpretations are left from the 
many that have been proposed during the decades since the invention of quantum
theory? I think, we do not have much of a choice at present%
\renewcommand{\thefootnote}{\fnsymbol{footnote}}%
\footnote[1]{The following  owes much to discussions with Dieter
Zeh, who finally convinced me that the Everett interpretation
{\em could} perhaps make sense at all.}%
\renewcommand{\thefootnote}{\arabic{footnote}}%
, {\em if} we
restrict ourselves to use only wavefunctions as kinematical
concepts (that is, we ignore hidden-variable theories, for example).

There seem to be only the two possibilities either (1) to
alter the Schr\"odin\-ger equation to get something like a ``real
collapse" \cite{GRW86,Pearle99}, or (2) to keep the theory unchanged and try to establish
some variant of the Everett interpretation.
Both approaches have their pros and cons, some of them are
listed in the following table.
%

%
Clearly collapse models face the immediate question of
how, when and where a collapse takes place. If a collapse
occurs before the information enters the consciousness of an
observer, one can maintain some kind of psycho-physical parallelism
by assuming that what is experienced subjectively is parallel to
the physical state of certain objects, e.g., parts of the brain. The 
last resort is to view consciousness as {\em causing} collapse, an
interpretation which can more or less be traced back to von Neumann.
In any case, the collapse happens with a certain probability (and with respect
to a certain basis in Hilbert space) and
this element of the theory comprises an {\em additional} axiom. 

How would we want to test such theories? One would look for collapse-like
deviations from the unitary Schr\"odinger dynamics. However, similar {\em apparent}
deviations are also produced by decoherence, in particular in the relevant
meso- and macroscopic range. So it is hard to discriminate these
{\em true} changes to the Schr\"odinger equation from the {\em apparent}
deviations brought about by decoherence\cite{Jo87}.

Everett interpretations lead into rather similar problems. Instead of 
specifying the collapse one has to define precisely how the wavefunction
is to be split up into branches. Decoherence can help here by selecting certain directions
in Hilbert space as dynamically stable (and others as extremely fragile
 -- branches with macroscopic objects in nonclassical
states immediately decohere), but the location of
the observer in the holistic quantum world is always a decisive ingredient. It must be assumed
that what is subjectively experienced is parallel to certain states
 (observer states)  in a certain
{\em component} of the global wave function. The probabilities (frequencies)
we observe in repeated measurements form also an additional axiom%
\begin{table}
\begin{tabular}{p{5.5cm}|p{5.5cm}l}
\makebox[5cm]{collapse models} & \makebox[5cm]{Everett} \\ \hline \hline
traditional psycho-physical parallelism: What is perceived is parallel
to the observer's physical {\em state} &
new form of psycho-physical parallelism: Subjective perception is
parallel to the observer state in a {\em component} of the universal wave function \\ \hline
probabilities put in by hand &
probabilities must also be postulated (existing ``derivations" are circular) \\
\hline
problems with relativity &
peaceful coexistence with relativity \\
\hline
experimental check: & experimental check:\\
look for collapse-like deviations from the Schr\"odinger equation &
look for macroscopic superpositions \\
\makebox[5cm]{$\Downarrow$} & \makebox[5cm]{$\Downarrow$} \\
hard to test because of decoherence & hard to test because of decoherence
\end{tabular}
\end{table}
\renewcommand{\thefootnote}{\fnsymbol{footnote}}%
\footnote[4]{There exist several claims in the literature that
probabilities can be derived in the Everett interpretation.
I think these proofs are circular. Consider a sequence of $N$ measurements
on copies of a two-state system, all prepared in the initial state
$a|1\rangle + b|2\rangle$. The resulting correlated state contains
$2^N$ components, where each pointer state shows one of the $2^N$
possible sequences of measurement results (e.g. as a computer printout). But these pointer states
are {\em always the same}, independently of the values of $a$ and $b$! Only
if each branch is given a {\em weight} involving $|a|^2$ and
$|b|^2$ one may recover the correct frequencies. See also \cite{Kent90}.
In addition, deep (and partially very old) questions about the meaning of
probabilities seem to reappear in the framework of Everett interpretations.}%
\renewcommand{\thefootnote}{\arabic{footnote}}%
.
The peaceful coexistence with relativity seems not to pose much
problems, since no collapse ever happens and all interactions are
local in (high-dimensional) configuration space.
But testing Everett means testing the Schr\"odinger equation in particular
with respect to macroscopic superpositions, and this again is precluded
by decoherence.

So it seems that both alternatives still have conceptual problems and
both are hard to test because of decoherence. We should not be surprised,
however, if it finally turned out that we do not know enough about consciousness
and its relation to the physical world to solve the quantum mystery
\cite{Squ90}.

%

\end{document}